\documentclass[twocolumn,twocolappendix, tighten]{aastex63}
\usepackage{graphicx}
\usepackage{url}
\usepackage{epstopdf}
\usepackage{color}
\usepackage{textcomp}
\usepackage{gensymb}
\usepackage{multirow}
\usepackage{ragged2e}
\usepackage{url}
\usepackage{amsmath} 
\usepackage{booktabs}
\usepackage{comment}
\usepackage{afterpage}
\usepackage{mathtools}
\usepackage{tabularx}
\usepackage{bold-extra}
\usepackage{xspace}
\usepackage{relsize}
\usepackage[utf8x]{inputenc} 
\usepackage{braket}

\usepackage[nolist,nohyperlinks]{acronym}

\usepackage{eht}

\usepackage[T1]{fontenc} 

\usepackage{comment}
\usepackage[encapsulated]{CJK}
\usepackage{ucs}
\usepackage[utf8x]{inputenc}
\newcommand{\cntext}[1]{\begin{CJK}{UTF8}{gbsn}#1\end{CJK}}

\DeclareGraphicsExtensions{.pdf,.png,.jpeg}

\begin{document}


\newcounter{iPap}\setcounter{iPap}{1}
\newcommand{\ehtsubtitle}{This is just the GAL for now}

\ifnum\value{iPap}=1 \renewcommand{\ehtsubtitle}{The Shadow of the Supermassive Black Hole in the Center of the Milky Way}\fi
\ifnum\value{iPap}=2 \renewcommand{\ehtsubtitle}{EHT and Multi-wavelength Observations, Data Processing, and Calibration}\fi
\ifnum\value{iPap}=3 \renewcommand{\ehtsubtitle}{Imaging of the Galactic Centre Supermassive Black Hole}\fi
\ifnum\value{iPap}=4 \renewcommand{\ehtsubtitle}{Variability, morphology, and black hole mass}\fi
\ifnum\value{iPap}=5 \renewcommand{\ehtsubtitle}{Testing Astrophysical Models of the Galactic Center Black Hole}\fi
\ifnum\value{iPap}=6 \renewcommand{\ehtsubtitle}{Testing the Black Hole Metric}\fi

\shorttitle{\ehtsubtitle}
\shortauthors{The EHT Collaboration et al.}

\title{
First Sagittarius A* Event Horizon Telescope Results.
\Roman{iPap}. \ehtsubtitle}

\author[0000-0002-9475-4254]{Kazunori Akiyama}
\affiliation{Massachusetts Institute of Technology Haystack Observatory, 99 Millstone Road, Westford, MA 01886, USA}
\affiliation{National Astronomical Observatory of Japan, 2-21-1 Osawa, Mitaka, Tokyo 181-8588, Japan}
\affiliation{Black Hole Initiative at Harvard University, 20 Garden Street, Cambridge, MA 02138, USA}

\author[0000-0002-9371-1033]{Antxon Alberdi}
\affiliation{Instituto de Astrof\'{\i}sica de Andaluc\'{\i}a-CSIC, Glorieta de la Astronom\'{\i}a s/n, E-18008 Granada, Spain}

\author{Walter Alef}
\affiliation{Max-Planck-Institut f\"ur Radioastronomie, Auf dem H\"ugel 69, D-53121 Bonn, Germany}

\author[0000-0001-6993-1696]{Juan Carlos Algaba}
\affiliation{Department of Physics, Faculty of Science, Universiti Malaya, 50603 Kuala Lumpur, Malaysia}

\author[0000-0003-3457-7660]{Richard Anantua}
\affiliation{Black Hole Initiative at Harvard University, 20 Garden Street, Cambridge, MA 02138, USA}
\affiliation{Center for Astrophysics $|$ Harvard \& Smithsonian, 60 Garden Street, Cambridge, MA 02138, USA}
\affiliation{Department of Physics \& Astronomy, The University of Texas at San Antonio, One UTSA Circle, San Antonio, TX 78249, USA}

\author[0000-0001-6988-8763]{Keiichi Asada}
\affiliation{Institute of Astronomy and Astrophysics, Academia Sinica, 11F of Astronomy-Mathematics Building, AS/NTU No. 1, Sec. 4, Roosevelt Rd, Taipei 10617, Taiwan, R.O.C.}

\author[0000-0002-2200-5393]{Rebecca Azulay}
\affiliation{Departament d'Astronomia i Astrof\'{\i}sica, Universitat de Val\`encia, C. Dr. Moliner 50, E-46100 Burjassot, Val\`encia, Spain}
\affiliation{Observatori Astronòmic, Universitat de Val\`encia, C. Catedr\'atico Jos\'e Beltr\'an 2, E-46980 Paterna, Val\`encia, Spain}
\affiliation{Max-Planck-Institut f\"ur Radioastronomie, Auf dem H\"ugel 69, D-53121 Bonn, Germany}

\author[0000-0002-7722-8412]{Uwe Bach}
\affiliation{Max-Planck-Institut f\"ur Radioastronomie, Auf dem H\"ugel 69, D-53121 Bonn, Germany}

\author[0000-0003-3090-3975]{Anne-Kathrin Baczko}
\affiliation{Max-Planck-Institut f\"ur Radioastronomie, Auf dem H\"ugel 69, D-53121 Bonn, Germany}

\author{David Ball}
\affiliation{Steward Observatory and Department of Astronomy, University of Arizona, 933 N. Cherry Ave., Tucson, AZ 85721, USA}

% ...

\collaboration{0}{The Event Horizon Telescope Collaboration}

\ifnum\value{iPap}=1 \author{Claudio Agurto}
\affiliation{European Southern Observatory, Alonso de C\'{o}rdova 3107, Vitacura, Casilla 19001, Santiago de Chile, Chile}

\author{Alexander Allardi}
\affiliation{University of Vermont, Burlington, VT 05405, U.S.A.}

\author{Rodrigo Amestica}
\affiliation{National Radio Astronomy Observatory, NRAO Technology Center, 1180 Boxwood Estate Road, Charlottesville, VA 22903, USA}

\author{Juan Pablo Araneda}
\affiliation{European Southern Observatory, Alonso de C\'{o}rdova 3107, Vitacura, Casilla 19001, Santiago de Chile, Chile}

\author{Oriel Arriagada}
\affiliation{European Southern Observatory, Alonso de C\'{o}rdova 3107, Vitacura, Casilla 19001, Santiago de Chile, Chile}

\author[0000-0003-2287-158X]{Jennie L. Berghuis}
\affiliation{Gemini Observatory, 670 N. A'ohoku Place, Hilo, Hawaii, 96720, USA}

\author{Alessandra Bertarini}
\affiliation{Max-Planck-Institut f\"ur Radioastronomie, Auf dem H\"ugel 69, D-53121 Bonn, Germany}
\affiliation{Institute of Geodesy and Geoinformation, University of Bonn, D-53113 Bonn, Germany}

\author{Ryan Berthold}
\affiliation{East Asian Observatory, 660 N. A'ohoku Place, Hilo, HI 96720, USA}

\author[0000-0002-2756-395X]{Jay Blanchard}
\affiliation{National Radio Astronomy Observatory, Socorro, NM 87801, USA}

\author{Ken Brown}
\affiliation{East Asian Observatory, 660 N. A'ohoku Place, Hilo, HI 96720, USA}

% ...

\fi 
\ifnum\value{iPap}=2 \include{./SAL2}\fi
\ifnum\value{iPap}=3 \include{./SAL3}\fi
\ifnum\value{iPap}=4 \include{./SAL4}\fi
\ifnum\value{iPap}=5 \include{./SAL5}\fi
\ifnum\value{iPap}=6 \include{./SAL6}\fi

\begin{acronym}
\acro{sn}[$S/N$]{signal-to-noise ratio}
\acro{vlbi}[VLBI]{very long baseline interferometry}
\acroplural{vlbi}[VLBI]{Very long baseline interferometry}
\acro{grmhd}[GRMHD]{general relativistic magnetohydrodynamics}
\acro{mhd}[MHD]{magnetohydrodynamics}
\acro{as}[as]{arcseconds}
\acro{agn}[AGN]{Active Galactic Nuclei}
\acro{llagn}[LLAGN]{low-luminosity AGN}
\acro{cena}[Cen~A]{Centaurus A}
\acro{sgra}[Sgr\,A*]{Sagittarius~A*}
\acro{agn}[AGN]{active galactic nuclei}
\acro{eht}[EHT]{Event Horizon Telescope}
\acro{bhc}[BHC]{\href{https://blackholecam.org}{BlackHoleCam}}
\acro{tanami}[TANAMI]{Tracking Active Galactic Nuclei with Austral Milliarcsecond Interferometry}
\acro{smbh}[SMBH]{supermassive black hole}
\acroplural{smbh}[SMBHs]{supermassive black holes}
\acro{aa}[ALMA]{Atacama Large Millimeter/submillimeter Array}
\acro{ap}[APEX]{Atacama Pathfinder Experiment}
\acro{pv}[PV]{IRAM~30\,m Telescope}
\acro{jc}[JCMT]{James Clerk Maxwell Telescope}
\acro{lm}[LMT]{Large Millimeter Telescope Alfonso Serrano}
\acro{sp}[SPT]{South Pole Telescope}
\acro{sm}[SMA]{Submillimeter Array}
\acro{az}[SMT]{Submillimeter Telescope}
\acro{chandra}[Chandra]{Chandra X-ray Observatory}
\acro{c}[c]{speed of light}
\acro{pc}[pc]{parsec}
\acro{gr}[GR]{general relativity}
\acro{aips}[\textsc{aips}]{\href{http://www.aips..edu}{Astronomical Image Processing System}}
\acro{casa}[CASA]{\href{https://casa..edu}{Common Astronomy Software Applications}}
\acro{symba}[\textsc{symba}]{\href{https://bitbucket.org/M_Janssen/symba}{SYnthetic Measurement creator for long Baseline Arrays}}
\acro{rpicard}[\textsc{Rpicard}]{\href{https://bitbucket.org/M_Janssen/picard}{Radboud PIpeline for the Calibration of high Angular Resolution Data}}
\acro{hops}[HOPS]{\href{https://www.haystack.mit.edu/tech/vlbi/hops.html}{Haystack Observatory Postprocessing System}}
\acro{hbt}[HBT]{Hanbury Brown and Twiss}
\acro{tov}[TOV]{Tolman-Oppenheimer-Volkoff}
\acro{mri}[MRI]{magnetorotational instability}
\acro{adaf}[ADAF]{advection-dominated accretion flow}
\acro{adios}[ADIOS]{adiabatic inflow-outflow solution}
\acro{cdaf}[CDAF]{convection-dominated accretion flow}
\acro{bz}[BZ]{Blandford-Znajek}
\acro{bp}[BP]{Blandford-Payne}
\acro{em}[EM]{electromagnetic}
\acro{blr}[BLR]{broad-line region}
\acro{nlr}[NLR]{narrow-line region}
\acro{ism}[ISM]{interstellar medium}
\acro{edf}[eDF]{electron distribution function}
\acro{pic}[PIC]{particle-in-cell}
\acro{sane}[SANE]{standard and normal evolution}
\acro{mad}[MAD]{magnetically arrested disk}
\acro{jive}[JIVE]{\href{http://www.jive.eu}{Joint Institute for VLBI ERIC}}
\acro{}[NRAO]{\href{https://www.nrao.edu}{National Radio Astronomy Observatory}}
\acro{muas}[$\mu$as]{microarcseconds}
\acroplural{agn}[AGN]{Active galactic nuclei}
\acro{jy}[Jy]{jansky}
\acro{pa}[PA]{position angle}
\acro{srmhd}[SRMHD]{special relativistic magnetohydrodynamics}
\acro{fov}[FOV]{field of view}
\acro{sefd}[SEFD]{system equivalent flux density}
\acroplural{sefd}[SEFDs]{system equivalent flux densities}
\end{acronym}


\begin{abstract}
We present the first Event Horizon Telescope (EHT) observations of Sagittarius~A* (\sgra), the Galactic center source associated with a supermassive black hole. These observations were conducted in 2017 using a global interferometric array of eight telescopes operating at a wavelength of $\lambda=1.3\,{\rm mm}$. The EHT data resolve a compact emission region with intrahour variability. A variety of imaging and modeling analyses all support an image that is dominated by a bright, thick ring with a diameter of $51.8 \pm 2.3$\,\uas (68\% credible interval). The ring has modest azimuthal brightness asymmetry and a comparatively dim interior. Using a large suite of numerical simulations, we demonstrate that the EHT images of \sgra are consistent with the expected appearance of a Kerr black hole with mass ${\sim}4 \times 10^6\,{\rm M}_\odot$, which is inferred to exist at this location based on previous infrared observations of individual stellar orbits as well as maser proper motion studies. Our model comparisons disfavor scenarios where the black hole is viewed at high inclination ($i > 50^\circ$), as well as non-spinning black holes and those with retrograde accretion disks. Our results provide direct evidence for the presence of a supermassive black hole at the center of the Milky Way galaxy, and for the first time we connect the predictions from dynamical measurements of stellar orbits on scales of $10^3-10^5$ gravitational radii to event horizon-scale images and variability.  Furthermore, a comparison with the EHT results for the supermassive black hole \m87 shows consistency with the predictions of general relativity spanning over three orders of magnitude in central mass.  
\end{abstract}


\nocite{PaperII}
\nocite{PaperIII}
\nocite{PaperIV}
\nocite{PaperV}
\nocite{PaperVI}

\section{Introduction}

Black holes are among the boldest and most profound predictions of Einstein's theory of General Relativity \citep[GR;][]{Einstein_1915}. 
Originally studied as a mathematical consequence of GR rather than as physically relevant objects \citep{Schwarzschild_1916}, they are now believed to be generic and sometimes inevitable outcomes of gravitational collapse \citep{Oppenheimer_1939,Penrose_1965}. In GR, the spacetime around astrophysical black holes is predicted to be uniquely described by the Kerr metric, which is entirely specified by the black hole’s mass and angular momentum or ``spin'' \citep{Kerr_1963}.

The first empirical evidence for their existence was through stellar-mass black holes, beginning with observations of X-ray binary orbits \citep{Bolton_1972,Webster_1972,McClintockRemillard_1986} and culminating in the detection of gravitational waves from merging stellar-mass black holes \citep{LIGO_2016}. In parallel, the discovery that quasars are not stellar in nature but are rather extremely luminous, compact objects located in the centers of distant galaxies  \citep{Schmidt_1963} led to an intensive effort to identify and measure the supermassive black holes (SMBHs) energetically favored to power them \citep{Lynden-Bell_1969}. Observations now suggest that SMBHs not only lie at the center of nearly every galaxy \citep{Richstone_1998} but also may play a role in their evolution \citep[see, e.g.,][]{Magorrian_1998,Fabian2012,Kormendy_2013}, though how exactly the ebbs and flows of black hole activity and growth proceed is a major outstanding question in the field.  

With the advent of the Event Horizon Telescope (EHT), SMBHs can now be studied with direct imaging \citep[][hereafter \m87~Paper~I-VIII]{M87PaperI,M87PaperII,M87PaperIII,M87PaperIV,M87PaperV,M87PaperVI,M87PaperVII,M87PaperVIII}. 
The combination of an event horizon and strong lensing near black holes is predicted to produce distinctive gravitational signatures in their images  \citep[e.g.,][]{Hilbert_1917,Bardeen_1973,Luminet_1979,Jaroszynski_1997,Falcke_2000}. In particular, simulated images of black holes typically have a central brightness depression encircled by a bright emission ring. The ring usually lies near the gravitationally lensed photon orbits that define the boundary of what we hereafter refer to as the black hole ``shadow.'' The shadow has an angular diameter $d_{\rm sh} \approx 10 G M/(c^2 D) \equiv 10 \theta_{\rm g}$, where $G$ is the gravitational constant, $c$ is the speed of light, $M$ is the black hole mass, and $D$ is the black hole distance.

From the first realization that SMBHs could power bright radio cores in many galactic nuclei \citep[][and references therein]{Lynden-Bell_1969}, the search has been on to identify them. Within our own galaxy, the compact source \sgra has been intensely studied as a candidate SMBH since its discovery as a bright source of radio emission located near the Galactic Center \citep{Balick_1974,Ekers_1975,Lo_1975}. Decades of monitoring its proper motion, as well as of motions of individual stars in orbit around it, have revealed \sgra to be an extremely dense concentration of mass ($M \approx 4\times10^6\msun$) that is located at and nearly motionless with respect to the dynamical center of the Galaxy ($D \approx 8\,{\rm kpc}$), providing strong evidence that it is the nuclear SMBH in our Galaxy \citep[e.g.,][]{Gravity_2019,Do_2019,Reid_2020}. As the nearest supermassive black hole, \sgra provides a unique opportunity to directly image such an object, together with its accretion system, in the most common, quiescent state of SMBHs across the Universe. It also provides the chance to elucidate some of the drivers of observed cycles in accretion power and jet launching, via comparison with the more ``active'' galactic nucleus \m87.

In this paper, we present the first EHT observations of \sgra and put them into context with our previous results on \m87. In \autoref{sec:comp}, we describe what was previously known about the physical properties of \sgra and compare them to \m87.  We then summarize our \sgra observations with the EHT and other observatories in \autoref{sec:obs} and discuss its variability in \autoref{sec:var}. In \autoref{sec:struct}, we present the first EHT images of \sgra and analyze its event-horizon-scale structure. In \autoref{ssec:astro}, we discuss the astrophysical interpretation of these results using an extensive suite of general relativistic magnetohydrodynamic (GRMHD) simulations, and in \autoref{ssec:GR} we present the constraints that these results give for GR and black hole alternatives. We provide the overall conclusions and outlook in \autoref{sec:discuss}. The five companion papers of this series provide a more comprehensive discussion of all these topics \citep[][hereafter Papers II-VI]{PaperII,PaperIII,PaperIV,PaperV,PaperVI}.

\section{Sgr A* and M87*}\label{sec:comp}

Decades of observations have provided a picture of our local SMBH that is unmatched in any other galaxy \citepalias[for details about the full spectrum, see][]{PaperII}. \sgra has been detected from long radio wavelengths ($\sim$1\,m) to the hard X-ray band, excepting approximately 1\,$\mu$m to 1\,nm due to extinction from dust in the galactic plane. \sgra is remarkable for its feeble emission, producing a bolometric luminosity of ${\lsim}10^{36}\,{\rm erg/s}$, only $\sim$100 times that of the Sun. Were it located in another galaxy, it would likely go undetected. Nevertheless, by observing its spectrum and variability, its environment, and its influence on surrounding bodies, a great deal has been learned about this source specifically, and about the astrophysical processes that operate around supermassive black holes. In this section we describe how we assembled our current knowledge of \sgra, discuss important theoretical uncertainties about its accretion and outflows, and compare it with the other horizon-scale EHT target, \m87.

\subsection{Properties of Sgr~A*}

The proximity of \sgra permits precise measurements of its gravitating mass via the monitoring of resolved individual stellar orbits. High-resolution infrared (IR) observations, using increasingly sophisticated instrumentation and analyses, have traced out the three dimensional orbits of several stars within the innermost arcsecond around \sgra \citep{Schodel_2002,Ghez_2003,Ghez_2008,Gillessen_2009,Gravity_2018,Do_2019,Gravity_2019}. These orbits jointly determine the mass and distance to \sgra to high precision, particularly the ratio $M/D$ that determines the angular size of the black hole on the sky. As discussed in \citetalias{PaperII} and \citetalias{PaperVI}, the current values for the mass and distance suggest an angular shadow diameter close to 50\,$\mu{\rm as}$, comparable to that of \m87. The closest orbital periapses confine the mass to within $\sim$1,000 Schwarzschild radii ($R_{\rm S}=2GM/c^2$).  

Radio observations of \sgra have provided a significant motivation for the development of the EHT experiment. \sgra shows a flat/inverted radio spectrum, which often arises from compact jet emission \citep{Blandford_Konigl_1979} in other low-luminosity active galactic nuclei (LLAGN; \citealt{Ho_1999,Nagar_2000}).  Such a spectrum can, however, result from any stratified, self-absorbed synchrotron source, where successively higher frequencies are produced at increasingly smaller scales, even without a jet \citep[e.g.,][]{Narayan_1995}. \sgra shows an excess of millimeter emission above the flat centimeter-wave spectrum, the so-called ``submillimeter bump,'' that was inferred to indicate the presence of a very compact emitting region at these wavelengths \citep[e.g.,][]{Zylka_1992,Falcke_1998}. 

To clarify the nature of this source, the structure of \sgra was investigated using very-long-baseline interferometry (VLBI) at progressively shorter wavelengths \citepalias[see][and references therein]{PaperII}. For wavelengths longer than several cm, the observed source size is entirely determined by scatter-broadening in the ionized interstellar medium, scaling with a wavelength dependence of $\lambda^2$. At wavelengths of 7~mm and shorter, the imprint of the intrinsic structure of \sgra became discernible through the scattering  \citep[e.g.,][]{Rogers_1994,Lo_1998,Doeleman_2001,Bower_2004,Shen_2005,Bower_2006}. The source grows more compact at shorter wavelengths, as expected \citep[e.g.,][]{Ozel_2000}, though it does not present a clear jet structure. For wavelengths as short as $\sim$1~mm, \sgra is only slightly blurred by scattering, and \citet{Falcke_2000} predicted that submillimeter VLBI could directly image a brightness depression within this region related to the black hole shadow. 

In parallel with these developments and motivated by the goal to study black holes, the capabilities of mm-VLBI improved rapidly. At 1.4\,mm, \citet{Padin_1990} reported the first VLBI fringes and \citet{Krichbaum_1998} obtained the first VLBI detections and associated source size measurements  for \sgra, each using a 2-element interferometer. \citet{Doeleman_2008} observed \sgra at 1.3\,mm with a 3-element VLBI array and reported the discovery of an intrinsic source size comparable to the expected angular diameter of the black hole shadow. These observations provided important constraints for theoretical models \citep[e.g.,][]{Broderick_2009,Dexter_2009,Moscibrodzka_2009} and strongly suggested that 1.3~mm VLBI has a clear view into the innermost region around the \sgra black hole. Subsequent VLBI studies of \sgra at 1.3\,mm with progressively enhanced arrays revealed the compact emission to be variable and significantly polarized, with measured ``closure phases'' that are indicative of persistent asymmetry in the image structure \citep{Fish_2011,Johnson_2015,Lu_2018}. 

Observations of \sgra outside the radio band were important for completing the picture of this source as an LLAGN.  When X-ray and gamma-ray instruments \textit{ROSAT} and \textit{Sigma/GRANAT} \citep{PredehlTruemper1994, Goldwurm_1994} could not identify a bright central source, it became clear that \sgra must either be obscured or anomalously faint compared to other known LLAGN.  The first identification of \sgra as a compact and variable source in the X-rays was achieved with the \textit{Chandra X-ray Observatory (Chandra)} \citep{Baganoffetal2001,Baganoffetal2003}. 
This detection, together with the ${\sim}4\times10^6$~M$_\odot$ mass of \sgra determined from stellar orbits, sets a maximum luminosity scale for the source, the so-called Eddington luminosity,\footnote{The Eddington luminosity is an idealized estimate of the maximum power for an accreting black hole: $L_{\rm Edd} \equiv 4\pi G M c m_{\rm p}/\sigma_{\rm T} \approx 10^{38}\left(M/M_\odot\right)$\,erg/s, where $\sigma_{\rm T}$ is the Thomson cross section and $m_{\rm p}$ is the mass of a proton.} and the X-ray measurements confirmed that \sgra is approximately 9 orders magnitude less luminous than its $L_{\rm Edd}$ --- the lowest Eddington ratio ($L/L_{\rm Edd})$ observed for any black hole.

This particularly faint high-energy emission stimulated a lively debate regarding the nature of \sgra's accretion/outflow properties, and motivated new theoretical developments.  If the emission primarily originates in the accretion inflow, either it must be exceptionally radiatively inefficient \citep{Narayan_1995}, or the accretion rate onto the black hole must be very small compared to what is captured, or possibly some combination of the two \citep{BlandfordBegelman1999, Narayan_2000, QuataertGruzinov2000, Yuan_Narayan_2014}. Alternately the emission could be dominated by an outflow, in which case a small accretion rate would also be favored \citep{Falcke_1993}, even during flares \citep{Markoff_2001}. 
Fortunately, the proximity of \sgra enables unparalleled study of the accretion flow. Observations with \textit{Chandra} marginally resolve thermal bremsstrahlung emission from near the gas capture radius, leading to an estimate of the captured accretion rate of ${\sim}10^{-6}-10^{-5} M_\odot$/yr at ${\sim}10^5R_{\rm S}$ \citep{Quataert2002, Baganoffetal2003}. The source of this gas can be connected to the observable stellar winds of ${\sim}30$ individual massive stars found in the inner parsec \citep{CokerMelia1999,RussellWangCuadra2017,Ressleretal2020}. The near-horizon accretion rate has also been estimated to be $10^{-9} - 10^{-7}$ $M_\odot$/yr through measurements of the Faraday rotation of polarized millimeter emission from \sgra \citep{Boweretal2003,Marrone_2006,Marroneetal2007}. This Faraday rotation is orders of magnitude smaller than would be expected if the accretion rate at the gas capture radius persisted to small radii \citep{Boweretal1999,Agol_2000,Quataert_pol_2000,Marrone_2006}.  This conclusion is further supported by radially resolved X-ray spectroscopy \citep{Wangetal2013} suggesting that only ${\sim}1\%$ of the captured mass makes it to the SMBH.  Taken together, the low luminosity, low radiative efficiency, and weak Faraday rotation are consistent with a weakly bound, magnetized accretion flow so diffuse that the electron and ion temperatures are unable to remain strongly coupled.

\begin{figure*}[ht]
\begin{center}
\includegraphics[width=0.49\textwidth]{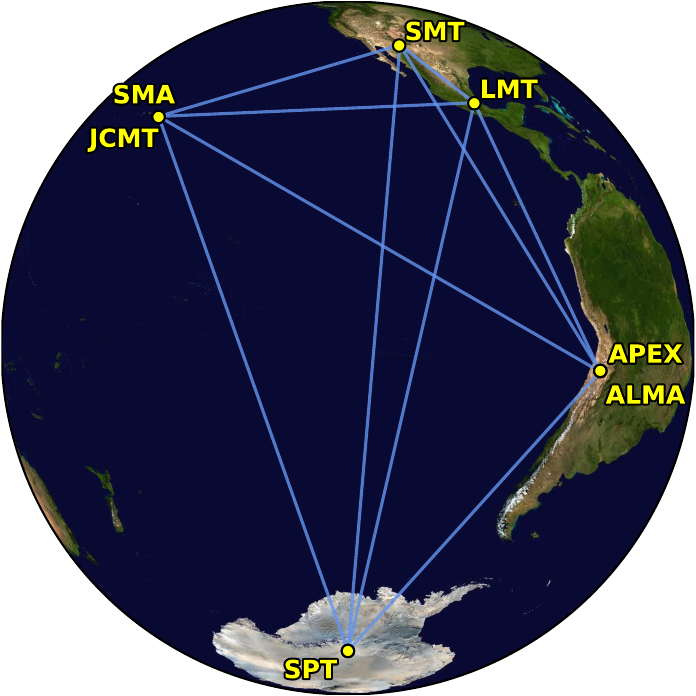}\hfill
\includegraphics[width=0.49\textwidth]{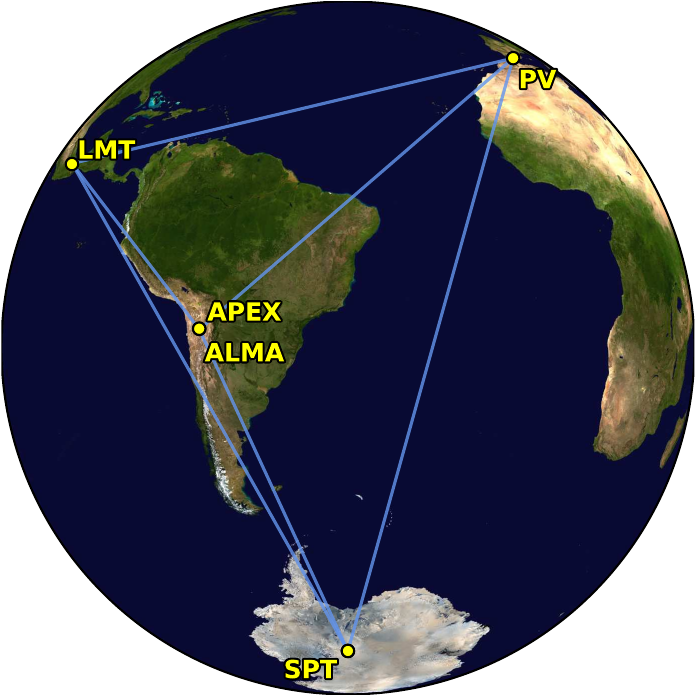}
\caption{The 2017 EHT array as seen from \sgra. The array included eight observatories at six locations: the \ac{aa} and the \ac{ap} on the Llano de Chajnantor in Chile, the \ac{lm} on Volc\'{a}n Sierra Negra in Mexico, the \ac{jc} and \ac{sm} on Maunakea in Hawai'i, the Institut de Radioastronomie Millim\'{e}trique 30-m telescope (PV) on Pico Veleta in Spain, the \ac{az} on Mt.\ Graham in Arizona, and the \ac{sp} in Antarctica. }
\label{fig:globe}
\end{center}
\end{figure*}

\begin{figure*}[t]
\begin{center}
\includegraphics[width=\textwidth]{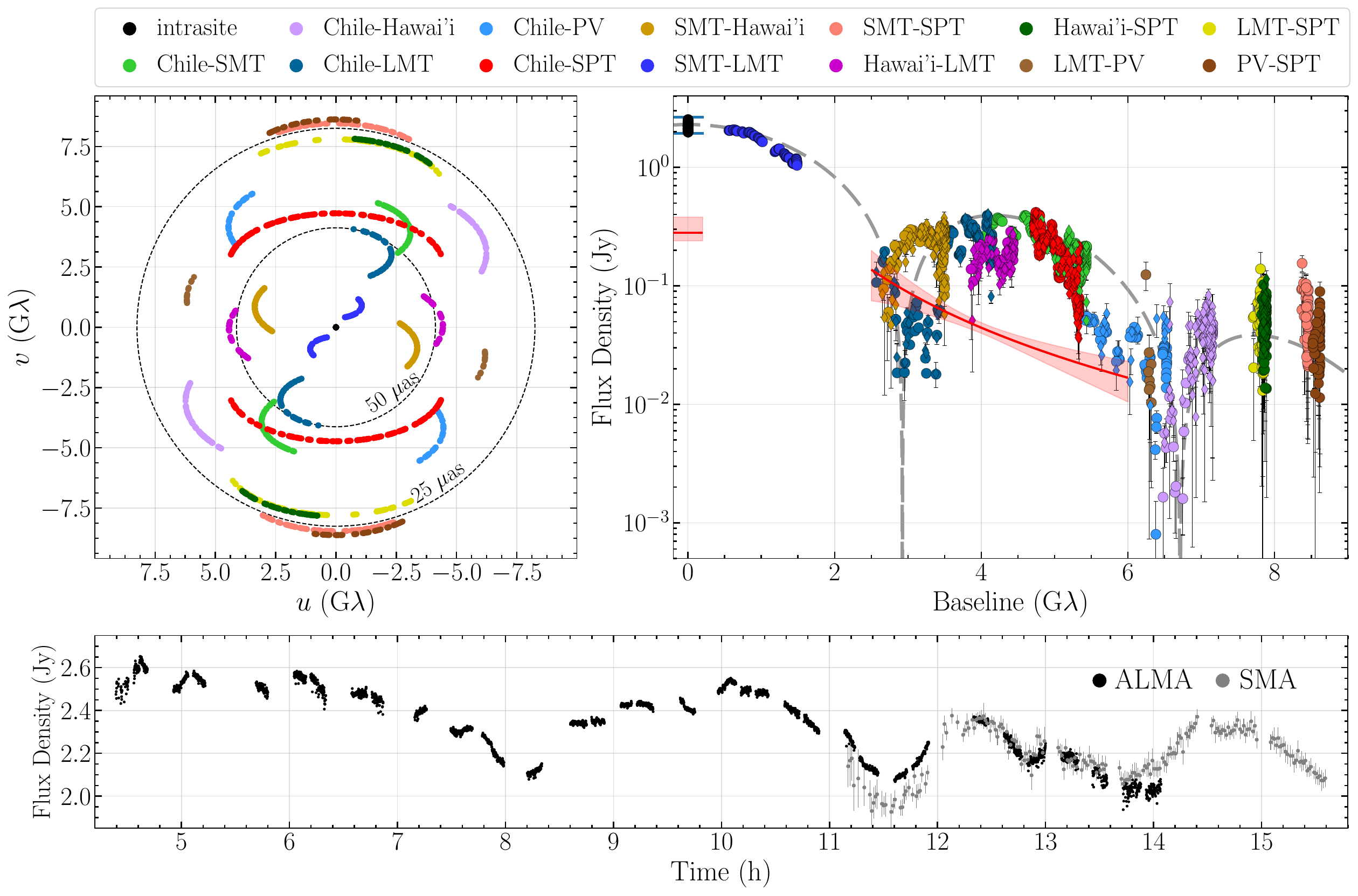}
\caption{
EHT observations of \sgra on April~7. 
Top-Left: EHT baseline coverage, where dimensionless coordinates $\vec{u} = (u, v)$ give the projected baseline vector for each antenna pair in units of the observing wavelength $\lambda$. 
Top-Right: Calibrated visibility amplitudes of \sgra as a function of projected baseline length $|\vec{u}|$. Error bars show ${\pm}1\sigma$ thermal (statistical) uncertainties. Diamonds denote baselines to APEX and JCMT to distinguish them from baselines to their co-located observatories, ALMA and SMA, respectively. The visibilities have been coherently averaged in 120~second intervals. For comparison, the gray dashed line shows the Fourier transform of a thin ring with diameter $54\,\mu{\rm as}$ that has been convolved with a circular Gaussian kernel of FWHM $23\,\mu{\rm as}$. The red line and shaded region show the root-mean-square variability and associated 68\%~credible interval over the range of baselines for which it can be accurately measured \citepalias[see][]{PaperIV}, while the blue horizontal ticks at zero baseline length show the range of variations in the total flux density.  Bottom: The full light curve of \sgra on April~7, measured using ALMA and the SMA as stand-alone interferometers.
}
\label{fig:vis}
\end{center}
\end{figure*}

Finally, \sgra exhibits flaring emission at most wavelengths, including continuous variability at millimeter wavelengths. 
In particular, approximately daily flares are observed at X-ray and NIR wavelengths that are often (but not always) simultaneous \citep[e.g.,][]{Eckart_2004} and that change on timescales as short as minutes \citep[e.g.][]{Baganoffetal2001,Genzel_2003}. These timescales suggest an origin from within ${\sim}5R_{\rm S}$ of the SMBH, which is consistent with astrometry of NIR flares using the GRAVITY instrument on the Very Large Telescope Interferometer \citep{Gravity_2018b}.  

\subsection{Comparison to \m87}
\m87 and \sgra have the largest angular sizes of any known SMBHs, making them the primary EHT targets. However, they differ in several important ways. First, they have substantially different luminosities and accretion rates, both in absolute terms and when scaled by mass. \m87, roughly 1500$\times$ more massive \citepalias[$M=(6.5 \pm 0.7) \times 10^9$\,\msun;][]{M87PaperI},   has an inferred accretion rate of $\dot{M} \sim 10^{-3}\,\msun/{\rm yr}$ and a bolometric luminosity of $L_{\rm bol} \approx 10^{42}\,{\rm erg/s}$ measured in 2017 \citep{EHT_MWL_2021}.  Estimates of the total kinetic power are typically $\sim10-100$ times larger, thus conservatively $L/L_{\rm Edd} \gtrsim 10^{-5}$ \citepalias{M87PaperVIII}.  Most likely this higher power indicates that \m87 is being fed directly from a larger reservoir of gas rather than a trickle from stellar winds.  This difference may underlie another substantial divergence between the sources: the prominent, powerful jet launched from \m87, which can be traced at multiple wavelengths and across nearly eight orders of magnitude in size \citep{EHT_MWL_2021}. The \m87 jet provides firm constraints on the source orientation with respect to the line of sight; thus, an inclination of $\sim20^\circ$ was used for numerical simulations in \citetalias{M87PaperV}, while for \sgra we do not have any such constraints.

The difference in the masses of \sgra and \m87 implies a similar difference in their variability timescales. The period of the innermost stable circular orbit (ISCO), which depends on the mass and spin of the black hole, serves as an approximate dynamical timescale. For prograde orbits, this period ranges from $4\pi t_{\rm g}$ (maximal spin) to $12\sqrt{6} \pi t_{\rm g}$ (zero spin), where $t_{\rm g} \equiv G M/c^3$. For \m87, this range corresponds to 5~days to 1~month, so the source structure is expected to be effectively unchanged over the course of an observing night. Indeed, EHT images of \m87 reconstructed on consecutive nights are almost identical \citepalias{M87PaperIV}. However, for \sgra, the range is only 4--30\,min, so the source structure can evolve within a single night.

\section{EHT Observations and Data Processing}\label{sec:obs}

In this section we summarize the EHT observations and data processing for \sgra. We refer the reader to \citetalias{M87PaperII} for a comprehensive discussion of the EHT instrument, \citetalias{M87PaperIII} for details on the 2017 observing campaign and data processing, and \citetalias{PaperII} for additional details specific to the \sgra data processing.

In 2017, the EHT observed \sgra\ on five nights between April 5 and 11 using an array with 8 participating observatories (see \autoref{fig:globe}).  Weather conditions were good or excellent at all sites on all five observing nights. The most sensitive element in the array, the Atacama Large Millimeter/submillimeter Array (ALMA), only observed \sgra on April 6, 7, and 11; the \ac{pv} only observed \sgra on April 7. In this initial series of papers, we focus on the observation with the best baseline coverage: April~7. In addition, we utilize the April~6 observations for testing, validation, and selected multi-day analyses. Our multi-wavelength coverage indicated that there was an X-ray flare on April~11,  accompanied by increased 1.3\,mm variability; we will consider that more complex data set in future work.  

Each site, except the JCMT and ALMA, received data in two circular polarizations simultaneously. The JCMT received a single circular polarization, and ALMA received orthogonal linear polarizations that are converted to a circular basis in post processing. All EHT sites recorded data in two frequency bands at 227.1 and 229.1~GHz, referred to as the ``low'' and ``high'' bands, respectively. The total recording rate at each fully-outfitted station is $32\,{\rm gigabits/second}$. The data were written to arrays of hard drives at each site, that were then brought from all sites to common locations where we computed the complex cross-correlation in the electric fields measured for each pair of stations. 

Following the initial computation of these correlations, residual phase and bandpass errors were corrected with two independent processing pipelines, EHT-HOPS \citep{Blackburn_2019} and rPICARD \citep{Janssen_2019}. The data were then \textit{a priori} calibrated using the system equivalent flux densities (SEFDs) for each telescope. Multiplication by the geometric mean of the SEFDs of the two stations on a given baseline converts dimensionless correlation coefficients to flux densities. SEFDs ranged from 60\,Jy at ALMA to $5\times10^4$\,Jy at low elevation at SMT. Further corrections are still required, as some sites do not measure the SEFD continuously, and in any case the SEFD does not capture many amplitude corrupting effects such as pointing and focus errors. A ``network calibration'' process, which uses redundancies in the array (e.g., co-located telescopes) to provide more accurate time-variable gain normalization for sites with co-located partners, was performed. \sgra presents a special case for calibration, as it varies significantly in time, and is surrounded by extended emission that corrupts the visibility amplitude for baselines within local arrays like ALMA and SMA. \citet{Wielgus_2022} discusses the techniques used to estimate the time-resolved flux density of \sgra for this calibration. For the remaining stations, gain corrections were computed using observations of the calibrator targets J1924-2914 and NRAO\,530 \citepalias{PaperII}.

\autoref{fig:vis} shows the EHT baseline coverage and interferometric visibility amplitudes (``visibilities'') for \sgra. The longest baselines have an interferometric fringe spacing of $1/|\vec{u}| \approx 24\,\mu{\rm as}$, which defines the diffraction-limited angular resolution of the EHT. The visibility amplitudes have two deep minima, the first at $|\vec{u}| \approx 3.0\,{\rm G}\lambda$ and the second at $|\vec{u}| \approx 6.5\,{\rm G}\lambda$. The amplitudes have a baseline dependence that is similar to that of an infinitesimally thin ring with a $54\,\mu{\rm as}$ diameter that has been blurred with a circular Gaussian kernel with $23\,\mu{\rm as}$ full width at half maximum (FWHM). The ring diameter is primarily constrained by the minima locations, while the width is determined by the amplitude of the secondary visibility peak between the minima. 
For instance, to be consistent with both a first minimum falling between $2.5-3.5\,{\rm G}\lambda$ and a second minimum falling between $6-7\,{\rm G}\lambda$, the ring must have a diameter of ${\sim}50-60\,\mu$as. 

However, the EHT data show evidence for complex and asymmetric source structure beyond a simple ring model, including a strong dependence on visibility amplitudes with the baseline position angle (especially near the first visibility minimum) and in the closure phases measured on triangles of EHT baselines, which differ significantly from $0^\circ$ and $180^\circ$. The EHT visibility amplitudes also show indications of residual calibration errors. To assess and quantify the source morphology of \sgra more generally, we applied a variety of imaging and modeling methods using both the interferometric visibility amplitude and phase information (\autoref{sec:struct}). These methods included a variety of approaches to account for residual calibration errors, including iterative self-calibration, simultaneous fitting of a model and the residual station gains, and analyses that used only closure quantities.

\section{Horizon-Scale Variability in Sgr~A*}\label{sec:var}

To characterize the spectrum and variability of \sgra, the EHT observing campaign was supported by parallel observations at several other observatories. Longer-wavelength VLBI observations, where interstellar scattering prevents direct observation of structural changes, were arranged to occur within a few days of the EHT campaign. IR and X-ray observations were arranged to be as simultaneous as possible with the EHT tracks on several days, and they resulted in one confirmed X-ray flare on April~11. 

For the observations of April~6 and 7, when no strong flares were observed in the parallel observations, we must evaluate whether \sgra has structural variations within our observation periods. The simplest evidence for variability at $\lambda=1.3\,{\rm mm}$ comes from analysis of the total flux density (the ``light curve''), which is measured during EHT observations using the ALMA and SMA connected element arrays. As discussed in \citet{Wielgus_2022}, the fractional variability across all days is approximately $9\%$, with variations of $4 - 13\%$ seen within individual nights (see, e.g., \autoref{fig:vis}). This variability is an order of magnitude stronger than is expected from interstellar scintillation. Thus, the light curves provide evidence for intrinsic image variability in \sgra. 

Each EHT baseline provides information about the detailed structure of this variability on an angular scale determined by the baseline length. On several triangles, EHT closure phases show slightly more variation than is expected from their thermal noise, the changing of projected baselines (from the source view) with Earth rotation, and interstellar scintillation \citepalias{PaperII}. Comparison of interferometric visibility amplitudes for nearby baselines also reveals variability that significantly exceeds what is expected from thermal noise, calibration uncertainties, and baseline evolution \citepalias{PaperIV}. 

To quantify the variability, we developed a simple parametric model for the spatio-temporal power spectrum for the variability of \sgra \citep{Georgiev_2022}. This model represents the variance in the visibility amplitude as a function of the radial distance in the $(u, v)$-plane, taking the form of a broken power law, as motivated by studies of GRMHD simulations. The source-integrated light curve is divided out in order to isolate structural variation from overall changes in flux density. The variability power spectrum was also empirically estimated by analyzing variations in visibility amplitudes on nearby baselines \citep{NoiseModeling}. This comparison included the same baselines sampled on different days, as well as nearby or crossing baseline tracks (e.g., SPT-LMT and SPT-SMA sample nearly identical baselines, but at different times). This analysis revealed that the fractional variability can be order unity for EHT baselines located near the two deep visibility minima (Figure~\ref{fig:vis}), even after normalizing the data to remove light curve fluctuations, and that it significantly exceeds the variability expected from interstellar scintillation \citepalias{PaperIV}.

\begin{figure}[t]
\begin{center}
\includegraphics[width=\columnwidth]{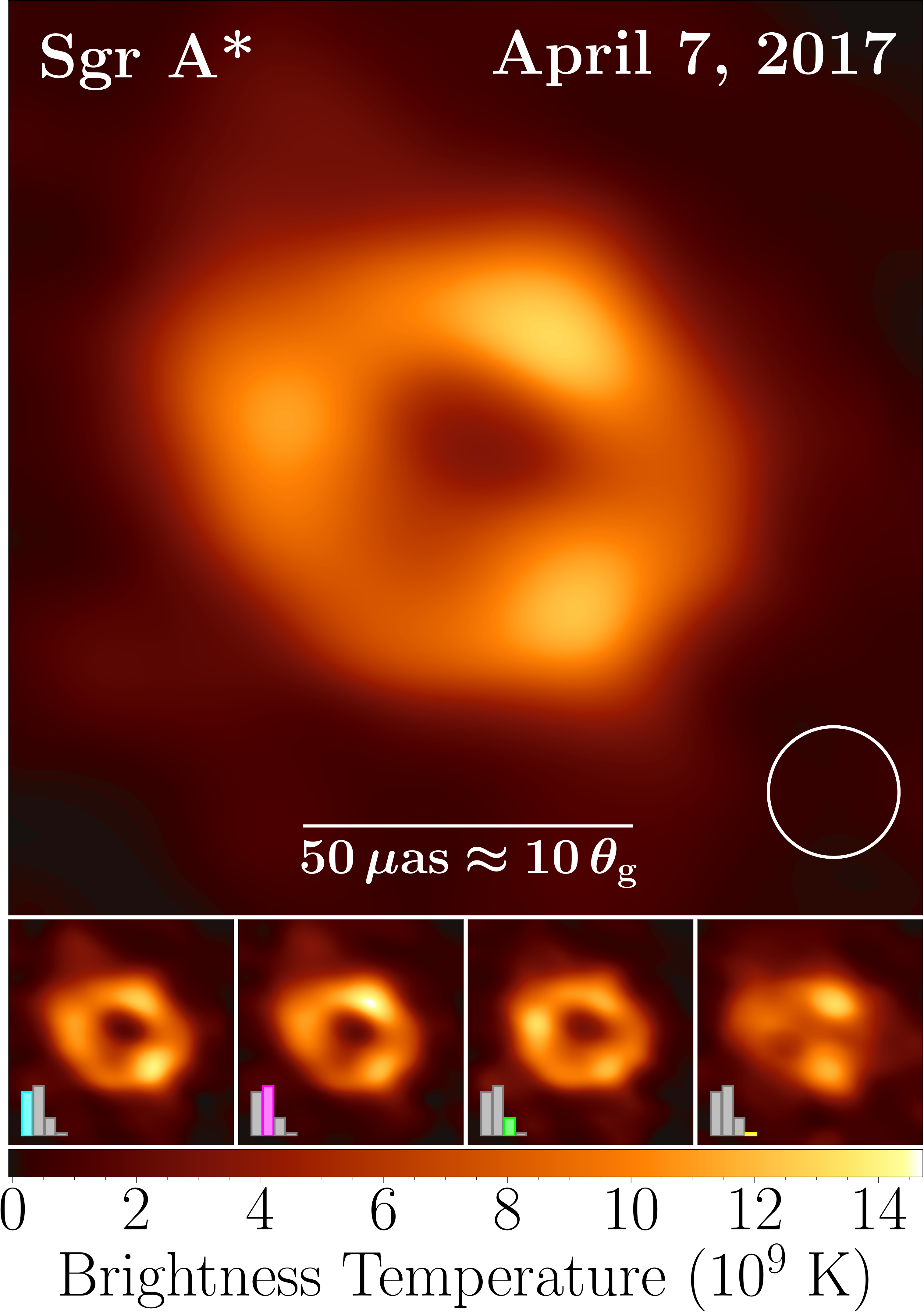}
\caption{
Representative EHT image of \sgra from observations on 2017 April 7. This image is an average over different reconstruction methodologies (CLEAN, RML, and Bayesian) and reconstructed morphologies. Color denotes the specific intensity, shown in units of brightness temperature. The inset circle shows the restoring beam used for CLEAN image reconstructions ($20\,\mu{\rm as}$ FWHM). The bottom panels show average images within subsets with similar morphologies, with their prevalence indicated by the inset bars. The multiplicity of image modes reflects uncertainty due to the sparse baseline coverage; it does not correspond to different snapshots of the variable source. Nearly all reconstructed images show a prominent ring morphology. While the diameter and thickness of the ring are generally consistent across the reconstructions, the azimuthal structure of the ring is poorly constrained. 
}
\label{fig:image}
\end{center}
\end{figure}

\section{Horizon-Scale Structure in Sgr~A*}\label{sec:struct}

Each interferometric visibility samples a single complex Fourier component of the image on the sky \citep[e.g.,][]{TMS}. Interferometric imaging algorithms seek to produce images from this sparse Fourier-domain information that are consistent with the data and physically plausible. Techniques such as the classical CLEAN algorithm and regularized maximum likelihood (RML) methods successfully produced EHT images of \m87, with remarkable agreement among methods \citepalias{M87PaperIV}. The EHT baseline coverage for \sgra is substantially better than for \m87, primarily because of the additional telescope (SPT) with mutual visibility of the source. Moreover, at $\lambda=1.3\,{\rm mm}$, \sgra has a compact flux density that is approximately four times larger than that of \m87, with no appreciable contribution to the short-baseline visibilities from an extended jet. However, producing an image of \sgra requires additional assumptions because of the rapid source variability and interstellar scattering. 

Specifically, VLBI imaging typically relies on Earth-rotation aperture synthesis, in which the projection of each baseline sweeps out an arc in the $(u, v)$-plane as the Earth rotates, allowing a sparse array of telescopes to obtain the $(u, v)$-coverage necessary for the imaging of a static source \citep{TMS}. To account for the source structural variability, we used a parametric model discussed in \autoref{sec:var}. By incorporating this variability error budget, imaging and modeling methods designed for a static source can be applied to analyze data from a variable source. 

To account for the interstellar scattering, we used two approaches \citepalias{PaperIII}. The first, ``on-sky imaging,'' applies no modifications to the data or images. In this approach, the algorithms simply reconstruct the scattered image of the source. The second, ``descattered imaging,'' adds an error budget to interferometric visibilities to account for stochastic scattering substructure before deconvolving the ensemble-average scattering kernel. Both the ensemble-average kernel and the power spectrum of scattering are used \citep{Psaltis_2018}, each of which is precisely known from an analysis combining of decades of observations of \sgra at centimeter wavelengths \citep{Johnson_2018}.

To test these imaging techniques and to select appropriate imaging parameters, we developed a suite of synthetic observations of seven geometric models that share the scattering and variability properties of \sgra. This suite included models with widely varying morphologies: rings, disks, a crescent, a double source, and a point-like source with an extended halo. Each model was selected to produce visibility amplitudes that were similar to those of \sgra, with two deep visibility minima, a physical scattering model applied, and stochastic temporal evolution generated by a statistical model \citep{Lee_2021}.

We then selected the sets of imaging parameters that accurately reconstruct images across the entire test suite, including both ring and non-ring data sets. These ``top set'' parameter choices yield a corresponding collection of reconstructed images of \sgra that provide both a representative average image and a measure of its uncertainty. In addition, we used a new Bayesian imaging method, which simultaneously estimates both the reconstructed image and its associated variability noise model \citep{Broderick_2020}. This method does not require training on synthetic data, although we used the same test suite for comparison and validation of this method. 

When applied to the \sgra data, over $95\%$ of the top set images have a prominent ring morphology. For an analysis using the combination of April~6 and 7 data, all samples from the Bayesian imaging posterior show a ring morphology. In addition, geometric modeling of the EHT data shows a consistent statistical preference for ring morphologies over alternatives with comparable complexity. The ring has a diameter, width, and central brightness depression that are consistent across the different choices of imaging methods and parameters \citepalias[see][]{PaperIII}. However, the reconstructed images show diversity in their specific attributes, particularly the azimuthal intensity distribution around the ring. This uncertainty is a consequence of the limited EHT baseline coverage, compounded by the challenges of imaging a variable source. We categorized the reconstructed images into four clusters spanning the primary reconstructed structures: three clusters are ring modes with varying position angle, while the fourth is a comparatively small set of reconstructed images with diverse non-ring morphologies. \autoref{fig:image} shows a representative average image of \sgra on April~7, as well as the average image for each of these clusters along with their relative occurrence frequency.  

To quantify the ring parameters in a complementary way, we used several geometrical modeling methods, the parametrization of which were guided by the reconstructed images of \sgra. These models are defined by a thick ring with azimuthal variations determined by low-order Fourier coefficients and an additional Gaussian brightness floor. Because these simple geometric models have a small number of parameters, they can be constrained using instantaneous snapshots of data. Hence, we used two modeling approaches. With ``snapshot'' modeling, we aggregate a series of independent fits to two-minute data segments. This approach does not require a variability model. With ``full-track'' modeling, we fit both a static geometric source model and a variability noise model simultaneously to the entire 12-hour observation. \autoref{tab:parameters} summarizes the consensus ring parameters measured using these methods \citepalias[for detailed results from individual methods, see][]{PaperIV}.

In \citetalias{PaperIII}, we also used dynamic imaging and snapshot ring modeling to analyze the intraday image variability in \sgra. We applied these analysis methods to the 100-minute intervals on April~6 and 7 with the best sampling \citep{Farah_2022}, adopting a strong ring prior to counteract the limited baseline coverage. On April~6, most dynamic imaging and modeling methods recover a nearly static image, while many reconstructions on April~7 find an evolving image. However, the results on April~7 are strongly affected by the underlying prior assumptions; different parameters in the dynamic imaging method result in different modes of position angle evolution in the reconstructed images, including some reconstructions that are nearly static. Thus, while the EHT data show detectable signs of image variability, we cannot reliably constrain the underlying image evolution. 

\begin{table}[t]
  \centering
    \caption{Measured Parameters of \sgra}
    \begin{tabular}{@{}lc}
        \toprule \toprule
        Parameter & EHT Estimate \\ 
        \midrule
        Emission ring:\footnote{
        The orientation and magnitude of the ring's brightness asymmetry is poorly constrained; it varies significantly among the reconstructed image modes and among different modeling and imaging methods. For details, see \citetalias{PaperIV}.\label{footIV}}\\
        \quad Diameter, $d$             & $51.8 \pm 2.3\,\mu{\rm as}$\\
        \quad Fractional width,         $W/d$    & ${\sim}\,30-50\%$\\
        \quad Orientation,              $\eta$  &---\\  
        \quad Brightness asymmetry,     $A$  & ${\sim}\,0.04-0.3$\\
        Angular gravitational radius,\textsuperscript{\ref{footIV}}   $\theta_{\rm g}$   & $4.8_{-0.7}^{+1.4}\,\mu{\rm as}$\\  
        Black hole mass,\footnote{To translate our estimate of $\theta_{\rm g}$ into an estimated mass of \sgra, we use the distance to \sgra estimated using trigonometric VLBI parallaxes and proper motions of molecular masers in spiral arms of the Milky Way \citep{Reid_2019}. For details, see \citetalias{PaperIV}.} $M$  & $4.0_{-0.6}^{+1.1}\times10^6 \,M_{\odot}$\\
        Angular shadow diameter,\footnote{Estimates of $d_{\rm sh}$ are determined solely from EHT data, but estimates of $\delta$ use priors for $\theta_{\rm g}$ from resolved stellar orbits as indicated. For details, see \citetalias{PaperVI}.\label{footVI}} $d_{\rm sh}$   & $48.7 \pm 7.0\,\mu{\rm as}$\\
        \multirow{2}{*}{Schwarzschild shadow deviation,\textsuperscript{\ref{footVI}} $\delta$} & $-0.08^{+0.09}_{-0.09}$ \text{(VLTI)}\\
         & $-0.04^{+0.09}_{-0.10}$ \text{(Keck)}\\
        \midrule
        Parameter & Previous Estimate \\         
        \midrule
        Angular gravitational radius, $\theta_{\rm g}$:\\
        \quad Stellar orbits (VLTI)\footnote{\citet{Gravity_2022}\label{footVLTI}}  & $5.125 \pm 0.009 \pm 0.020\,\mu{\rm as}$ \\          
        \quad Stellar orbits (Keck)\footnote{\citet{Do_2019}\label{footKeck}}  & $4.92 \pm 0.03 \pm 0.01\,\mu{\rm as}$  \\          
        Black hole distance, $D$:\\
        \quad Stellar orbits (VLTI)\textsuperscript{\ref{footVLTI}}  & $8277 \pm 9 \pm 33\,{\rm pc}$ \\          
        \quad Stellar orbits (Keck)\textsuperscript{\ref{footKeck}}  & $7935 \pm 50 \pm 32\,{\rm pc}$ \\          
        \quad Masers
      (cm~VLBI)\footnote{\citet{Reid_2019}\label{footMaser}
      \\{\hspace*{-\parindent}\parbox{\columnwidth}{\normalsize
      \rule[5pt]{0pt}{10pt}
      {{\textbf{Notes.}} Stated uncertainties correspond to 68\% credible intervals.}  }}}  & $8150 \pm 150\,{\rm pc}$ \\          
        Black hole mass, $M$:\\
        \quad Stellar orbits (VLTI)\textsuperscript{\ref{footVLTI}}  & $(4.297 \pm 0.013) \times 10^6 M_\odot$ \\
        \quad Stellar orbits (Keck)\textsuperscript{\ref{footKeck}}  & $(3.951 \pm 0.047) \times 10^6 M_\odot$ \\
        \bottomrule
    \end{tabular}
    \label{tab:parameters}
\end{table}

\section{Implications for Accretion and Outflow Physics}\label{ssec:astro}

What can we learn from these images and their variability properties?  Focusing first on the astrophysics of the accretion process and jet launching, we can explore which physical scenarios are most consistent with our results, under the assumption that Sgr A* is a Kerr black hole with mass and distance accurately determined from stellar orbits. Our constraints on the properties of the black hole and potential deviations from GR are explored in Section~\ref{ssec:GR}. 

We assume that the accretion structure around \sgra is approximately governed by ideal GRMHD, as was done for \m87, which is common in the literature for modeling SMBHs \citep[see, e.g.,][]{Gammie_2003}.  Decades of observations and semi-analytical modeling (see \citetalias{PaperV} for details, references, and caveats) constrain the average plasma properties close to the event horizon of \sgra, allowing us to make several additional simplifying approximations. In particular, for \sgra we can assume that radiative cooling does not strongly affect the dynamics, and that the electrons and ions are weakly coupled by Coulomb collisions, so that ions and electrons can have distinct temperatures in some parts of the flow.  

Because we model the plasma as a fluid with a single temperature, one of our main sources of uncertainty is the treatment of the electrons, whose presence is not explicitly accounted for in the simulation evolution equations. We explore several parameterized models to assign the electron distribution function (eDF), assuming that the electron temperature is proportional to the proton temperature, with a proportionality that depends on the gas-to-magnetic pressure ratio \citep{Chan_2015}. The eDFs include thermal and variations of non-thermal, the latter of which were not explored in the \m87 2017 papers.  Our fiducial thermal models employ the same eDF prescription as for the \m87 papers, using only one free parameter, $\Rh$, to specify the proton-to-electron temperature ratio in regions where gas pressure dominates the magnetic pressure \citep{Moscibrodzka:2016}. This ratio is typically larger in the disk midplane than in the jet/outflow. Since the radiation is produced by electrons, increasing $\Rh$ effectively increases the brightness of the jets/outflow region relative to the disk, and changes the resulting images/spectra \citepalias{M87PaperV}.  Compared to \m87, we also allow the inclination angle to vary.  

We employ five different ideal GRMHD codes to explore a large swath of overlapping parameter space, in some cases with very similar set-ups allowing for consistency checks.  In other cases, with a more exploratory sampling of parameter space, we also allow differences in, for example, adiabatic index, resolution, and size of the tori and/or computational domain.  Most models are initialized with an orbiting torus of plasma with characteristic radius $\sim 20 G M/c^2$.  The torus is seeded with a weak, poloidal magnetic field, and can be either prograde or retrograde with respect to the black hole spin (a free parameter). We also consider a limited set of exploratory models such as those with `tilt' where the black hole spin axis is misaligned from the rotation axis of the torus \citep{Liska_2018,Chatterjee_2020,WhiteQuataert_2022}, as well as a model initiated on a very large grid using a more realistic set-up for the outer boundary conditions in which the accretion flow is directly fed by winds from orbiting stars \citep{Ressleretal2020}. Both of these are more realistic physical scenarios but also allow a much larger range of parameter space than we could fully explore.  

We classify the fiducial models as being in either the magnetically arrested  \citep[MAD;][]{Narayan_2003} or standard and normal evolution \citep[SANE;][]{Narayan_2012} modes. In MAD models the ordered magnetic fields significantly affect the dynamics of the flow, episodically halting accretion onto the black hole, while SANE models have weaker, more turbulent magnetic fields. Because the dynamical timescale in \sgra is short compared to a night of observations, it is important to run each model for enough time to capture the range of spectral and structural variations. The simulations are typically run for $30{,}000 t_{\rm g}$, while some are run for more than $100{,}000 t_{\rm g}$ in order to sample a broader distribution of behavior.

\begin{figure*}[t]
\begin{center}
\includegraphics[width=\textwidth]{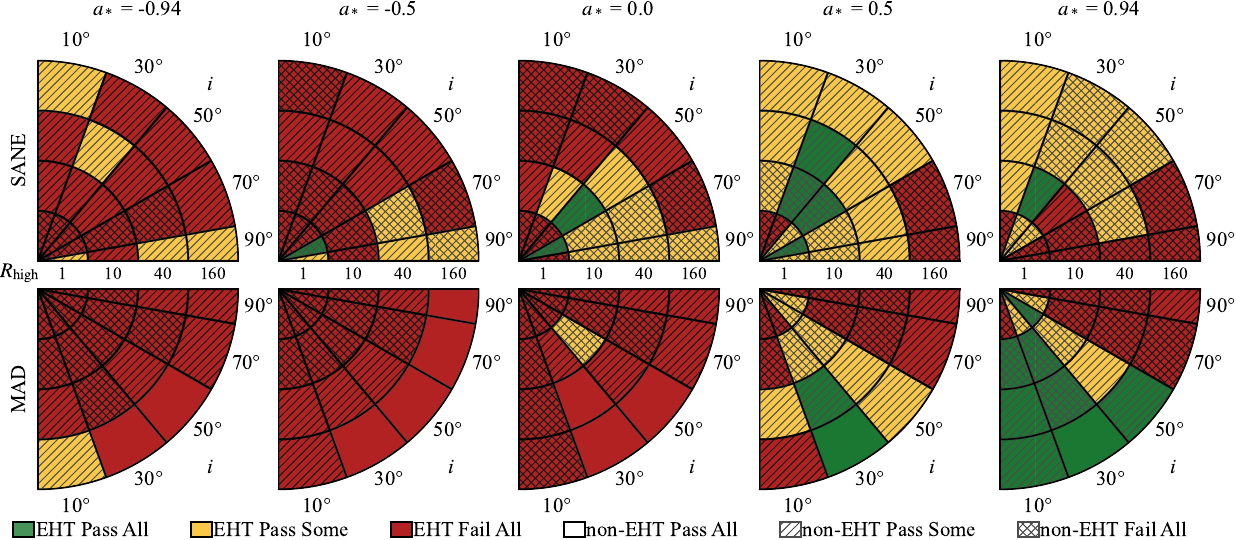}
\caption{
Summary of constraints on our 200 fiducial GRMHD simulations. Color indicates combined EHT constraints apart from structural or flux variability, and hatching indicates combined non-EHT constraints. For each constraint category and parameter combination, we delineate whether all of the three simulation codes run with those parameters pass, whether only some pass, or whether none pass. These exclusions leave only two models, each a MAD with prograde spin, $30^\circ$ inclination, and $\Rh=160$. For details, see \citetalias{PaperV}.
}
\label{fig:grmhd_constraints}
\end{center}
\end{figure*}

For each time-dependent GRMHD simulation, with an eDF prescription and inclination with respect to the line-of-sight, we calculate a sequence of model images (movies) using ray tracing and including synchrotron emission and absorption.  We also calculate spectra including synchrotron emission and absorption, bremsstrahlung emission and, using Monte Carlo methods,  Compton scattering.   These synthetic data sets are then used to generate simulated EHT images as well as multi-wavelength light curves and spectral energy distributions (SEDs) for comparison with the \sgra data described in \citetalias{PaperII}.  We scale all images to a benchmark average flux density of 2.4\,Jy at 230\,GHz to match the average synchrotron flux density of \sgra (see \citetalias{PaperII} and \citealt{Wielgus_2022}). 

We evaluate the simulations against three types of observational constraints: EHT interferometric measurements, emission at other wavelengths, and variability. The EHT constraints include 1) a measure of the image size, 2) salient features from the visibility amplitudes, such as the location of the first deep minimum, and 3) the diameter, asymmetry, and width of simplified ring models fitted to well-sampled portions of the April~7 visibility data. The constraints from other wavelengths include the flux densities at 86\,GHz, $2.2\,\mu{\rm m}$ in the NIR, and X-ray, and the major axis source size at 86\,GHz, constrained from observations with the Global mm-VLBI Array. Finally, the variability constraints are 1) the fractional 230\,GHz variability on 3\,h timescales, derived from more than a decade of measurements, and 2) the structural variability of the source, calculated at a baseline length of 4G$\lambda$ after fitting a parameterized model to the visibility amplitude variation versus baseline length. See \citetalias{PaperV} for the full ranges of tests and pass/fail conditions. 

Compared to \citetalias{M87PaperV}, we explore a larger range of models and model parameter space, and we also include some additional observational constraints. These include the degree of intrinsic variability as well as the broadband spectral constraints given above. 
Accordingly, we find that all our models fail at least one of the observational constraints. These results indicate the power of combining interferometric data with other observational constraints to narrow down the viable physical parameter space. We now summarize our main results and their implications for our understanding of \sgra's accretion state and geometry. 

We primarily focus on a set of ``fiducial'' simulations, which use aligned (prograde or retrograde) accretion flows and thermal eDFs defined via the $R_{\rm high}$ prescription. We declare a model to fully pass a set of constraints only when all GRMHD simulations with those parameters pass. This approach helps to ensure that our selection of favored models is resilient to small variations in the GRMHD simulation choices and software. \autoref{fig:grmhd_constraints} summarizes these results.

All edge-on (high inclination) models fail the combined set of EHT-only constraints for at least one simulation, and almost all retrograde models ($\abh < 0$) fail. There are two interesting groupings of models that pass all EHT constraints for all simulations:  both have positive/prograde spin ($\abh=0.5$, 0.94) with lower (${\le}\,50^\circ$) inclination, but some are MAD (10) and some are SANE (8).  With only ${\sim}10\%$ pass rate for all models, it is clear that EHT imaging data alone are capable of strongly down-selecting the potential model space. The more heterogeneous non-EHT constraints prefer a rather different set of models.  However even with all these constraints, 11/200 models pass for all simulations; all of these are MADs with all but one having $R_{\rm high} = 160$.  On the other hand, they cover a wider range of spin compared to EHT-only constraints, with a slight preference for $\abh \le 0$ and higher inclinations, and including retrograde and edge-on models.

\begin{figure*}[t]
\begin{center}
\includegraphics[width=\textwidth]{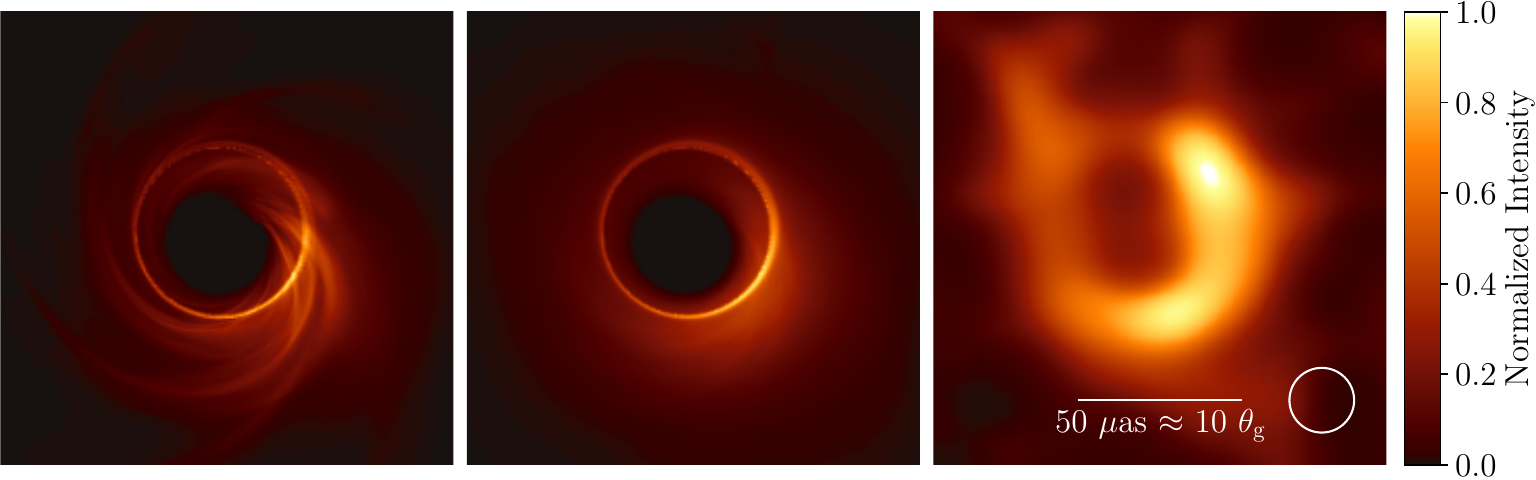}
\caption{
Simulated images of \sgra. Left: A single snapshot image of a numerical simulation of \sgra that passes 10 out of the 11 observational criteria described in \citetalias{PaperV}. 
Middle: The average of this simulation with time sampling that matches the EHT observational cadence on April~7.
Right: Representative image reconstruction using synthetic visibilities generated from the simulation in the adjacent panels \citepalias[see Appendix~H in][]{PaperIII}. This image has been averaged across methodologies and reconstructed morphologies, as in \autoref{fig:image}. Each panel is shown on a linear brightness scale that is normalized to its peak.
}
\label{fig:grmhd}
\end{center}
\end{figure*}

None of the 200 fiducial models pass all 11 constraints in combination, the most strict of which are the 86\,GHz size, the ring diameter and the light curve variability on 3-hour timescales.  Of these, the light curve variability turns out to be the most stringent constraint, passing only 4\% of fiducial models. SANE models, which are less variable than MADs, are preferred, while all the other EHT/non-EHT constraints generally favor MADs. If we consider the set of models that pass 10/11 constraints, it is notable that for both MADs and SANEs, there is still a marked preference for models with prograde spin, a lower inclination and $\Rh>40$ with several clustered at the maximum of $\Rh=160$. 

The ``exploratory'' models cover less of a range in parameter space but still indicate important trends.  
In particular, including non-thermal electrons (which are especially important for modeling flares) tends to push the limits of allowed NIR flux and, to some extent, also the X-ray flux. Because the non-thermal particles also enhance the 230\,GHz flux density, rescaling to a fixed flux density results in a smaller accretion rate. The smaller resulting opacity then affects the image properties, such as producing a somewhat narrower ring morphology at 230\,GHz and a larger image at 86\,GHz. Nevertheless, the addition of non-thermal models does not drastically change the preferred parameter space, and passing models also favor prograde spin and lower inclination. Increased tilt of the accretion flow tends to increase variability and the NIR flux, and thus leads to model failures. Similarly, neither of the two wind-fed models passed, but this is not surprising as they only model a single spin ($\abh=0$) and two instances of the thermal $\Rh$ eDF.  With such sparse sampling of parameter space, these classes of models require a more focused study to draw conclusions.

Overall, very few of our models are as quiet as the data. Although this wasn't investigated in \citetalias{M87PaperV}, subsequent work suggests a similar result for \m87 simulations \citep{Satapathy_2022}. In general SANE models are less variable than MADs, and face-on models are somewhat less variable than edge-on. However, there are limitations of the modeling that may affect the variability. For instance, collisionless effects, radiative cooling, and improved electron heating models could all potentially reduce variability. Thus, our variability constraints may be too strict. 

There are only two models that pass everything but the variability constraints. The models have similar parameters: both are MAD, both are prograde spin (one has $\abh=0.5$, and the other has $\abh=0.94$), and both have $i=30^\circ$ and $\Rh=160$. \autoref{fig:grmhd} shows a representative snapshot from one of these simulations and its corresponding image reconstruction. If we take these two models as indicative, there is a preference for accretion rates at the lower end of the range found by other constraints, $10^{-9}-10^{-8}\msun/{\rm yr}$, and an outflow power of ${\sim}10^{38}$\,erg/s, which is comparable to the bolometric luminosity of stellar-mass black holes in X-ray binaries. The combination of high $\Rh$ and relatively low inclination satisfies the spectral constraints without their associated jet structure producing too much asymmetry to satisfy the EHT constraints. A low inclination is also consistent with the independent estimate based on tracking the motions of NIR flaring structures using the VLTI \citep{Gravity_2018b}. 
Our results highlight the value of continued EHT plus multiwavelength monitoring of \sgra, and there is clearly much more exploration to conduct, including improved theoretical models and numerical simulations, full SED modeling, and polarimetric imaging, all of which will yield deeper physical insights \citep[see, e.g.,][]{M87PaperVIII}.

\section{Implications for Black Holes and General Relativity}\label{ssec:GR}

As demonstrated using the suite of GRMHD simulations discussed in \autoref{ssec:astro}, the EHT images of \sgra are consistent with the expected appearance of a Kerr black hole. Moreover, our images are consistent with an angular gravitational radius $\theta_{\rm g}$ that matches the expectations from nearly Keplerian orbits of stars on scales of $(10^3 - 10^5) R_{\rm S}$. 

To quantify this consistency, we compute $\theta_{\rm g}$ using EHT measurements alone \citepalias[for details, see][]{PaperIV}. Following the procedure developed in \citetalias{M87PaperVI}, we estimate $\theta_{\rm g}$ by calibrating the observed emission ring diameter $d$ to the known angular gravitational scale $\theta_{\rm g}$ for a suite of synthetic datasets produced from the GRMHD library presented in \citetalias{PaperV}. Specifically, we generate 100 datasets from GRMHD simulations with varying intrinsic $\theta_{\rm g}$ and that span the explored range in accretion states, black hole spins, viewing inclinations, position angles, and phenomenological electron heating parameters $\Rh$. 
Each synthetic dataset matches the baseline coverage and sensitivity of the EHT observations. We then estimate the ring diameter using multiple geometric modeling and imaging methods, deriving a separate calibration factor $\alpha \equiv d/\theta_{\rm g}$ for each method and for every dataset to attain a method-dependent scaling relationship and associated uncertainty. 

With this approach, we estimate $\theta_{\rm g} = 4.8_{-0.7}^{+1.4}\,\mu{\rm as}$ for \sgra, where the uncertainties correspond to a 68\%~credible  interval. This estimate is consistent with, but much less constraining than, measurements of $\theta_{\rm g}$ using resolved stellar orbits (see \autoref{tab:parameters}). This procedure assumes the validity of the Kerr metric and relies on our suite of GRMHD simulations to provide a reasonable proxy for the space of viable emission models. The fractional uncertainty in our estimate of $\theta_{\rm g}$ for \sgra is broader than our estimate for \m87 ($\theta_{\rm g} = 3.8 \pm 0.4\,\mu{\rm as}$), primarily because of the increased calibration uncertainty associated with the unknown inclination of \sgra and because of the increased diameter measurement uncertainty associated with intrinsic variability of \sgra.

We also used EHT images to constrain potential deviations from the Kerr metric and to test the nature of the compact object in \sgra \citepalias[for details, see][]{PaperVI}. For instance, the brightness depression has a contrast $f_{\rm c} \lsim 0.3$, which provides support for the existence of an event horizon in \sgra. If the compact object instead had an absorptive boundary that radiated the thermalized  energy of infalling material $\dot{M} c^2$, it could still produce a depression in the EHT images but would generate IR emission that greatly exceeds the measured spectrum of \sgra \citep{Broderick_2006,Narayan_2008}. Alternatively, a partially reflecting surface would reduce the depth of the depression; the EHT images directly constrain the albedo of such a reflecting surface to be ${\lsim}\,0.3$. The brightness depression also rules out several specific black hole alternatives, including some models of naked singularities \citep[e.g.,][]{Joshi_2014} and some models of boson stars \citep[e.g.,][]{Olivares_2020}, although other horizonless black-hole mimickers produce apparent shadows that are consistent with the observed depression \cite[e.g.,][]{Shaikh_2019}.

The measured ring diameter also provides constraints on the spacetime metric. To obtain these constraints, we generate a series of synthetic images to relate a measured emission ring diameter to that of the underlying black hole shadow. These images included a broad range of GRMHD simulations from \citetalias{PaperV}, as well as images for which the GRMHD and underlying metric assumptions are relaxed. Specifically, we generated sets of images of analytic models for accretion onto 1) a Kerr black hole, 2) a black hole with parametric deviations from Kerr given by the Johannsen-Psaltis metric \citep{Johannsen_2011}, and 3) a non-Kerr black hole defined by the Kerr-Sen metric \citep{Garcia_1995}. For each analytic model, we allow the emission prescriptions to vary within physically plausible limits \citep[for details, see][]{Ozel_2021,Younsi_2021}. We find a similar relationship between the diameter of the emission ring and that of the black hole shadow in all these cases, indicating that this relationship is insensitive to the details of the underlying spacetime.

Selecting a subset of these models, we then generate 145 synthetic datasets.  We apply imaging and modeling methods to compute the ring diameter for each dataset to evaluate method-dependent measurement uncertainties and biases. We then use this calibration together with the EHT measurements of the ring diameter to determine the angular diameter of the black hole shadow for \sgra: $d_{\rm sh} = 48.7 \pm 7.0\,\mu{\rm as}$. This result is tighter than the range for $\theta_{\rm g}$ derived above because of differences in the calibration datasets and procedures used in \citetalias{PaperIV} and \citetalias{PaperVI}. Specifically, \citetalias{PaperIV} uses a calibration suite that consists entirely of dynamic GRMHD models, and it derives a scale factor between $\theta_{\rm g}$ and the angular diameter of the emission ring. In contrast, \citetalias{PaperVI} uses a calibration suite containing both static Kerr and non-Kerr images and dynamic GRMHD models, and it derives a scale factor between $d_{\rm sh}$ and the angular diameter of the emission ring. \citetalias{PaperVI} also excludes datasets for which the range of reconstructed diameters is more than $2-3$ times the range that is measured using the \sgra data. Thus, comparison of these results provides a measure of the impact of our assumptions and procedures on the inference of the black hole properties.

By comparing this shadow diameter with stellar dynamical measurements of the mass of \sgra, we also determine a deviation parameter $\delta$, which quantifies the fractional difference between the inferred shadow diameter and its expected value for a non-spinning (Schwarzschild) black hole \citepalias[for details, see][]{PaperVI}. We find $\delta = -0.08^{+0.09}_{-0.09}$ when using VLTI measurements of $\theta_{\rm g}$ and $\delta = -0.04^{+0.09}_{-0.10}$ when using Keck measurements of $\theta_{\rm g}$ (see \autoref{tab:parameters}). For comparison, a spinning (Kerr) black hole has $-0.08 \leq \delta \leq 0$. 

\begin{figure}[t]
\begin{center}
\includegraphics[width=\columnwidth]{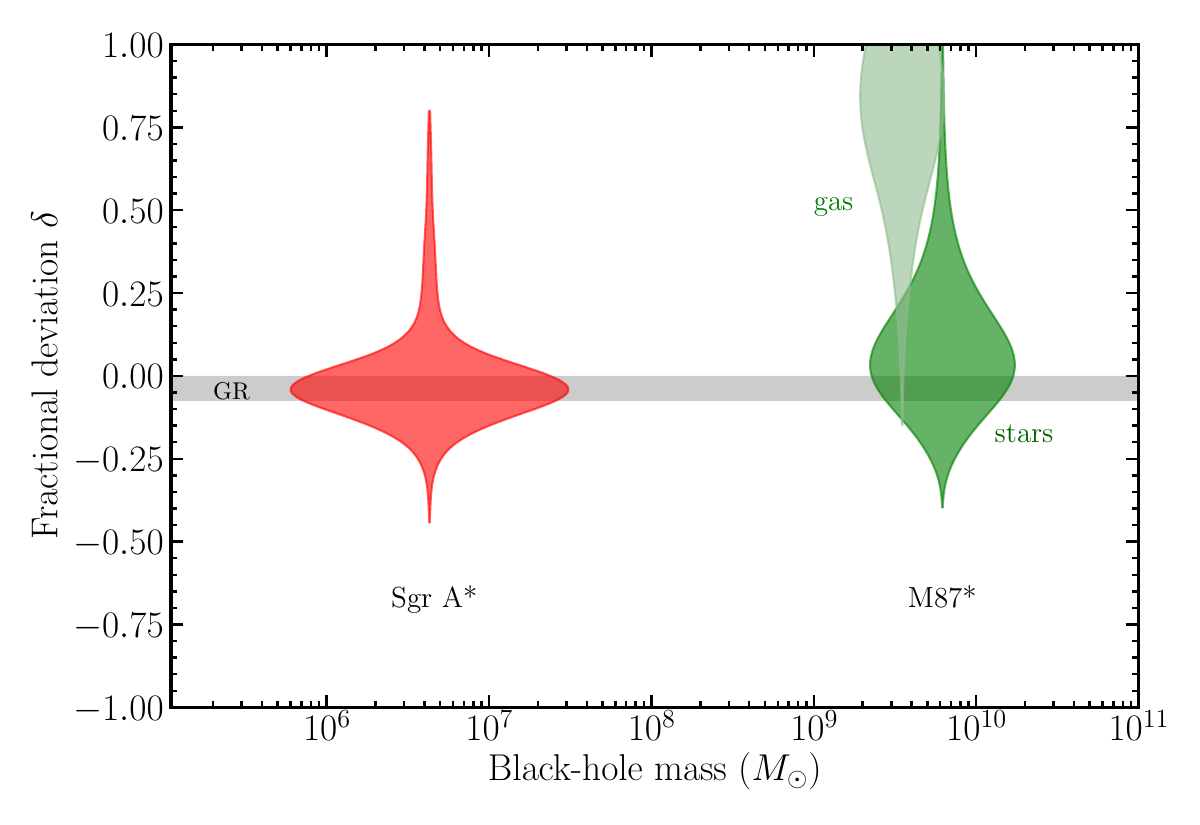}
\caption{
Comparison of posterior distributions for the Schwarzschild fractional shadow deviation parameter $\delta$ measured by the EHT for \sgra and \m87. For \sgra, $\delta$ is computed relative the expected shadow size from monitoring stellar orbits; for \m87, $\delta$ is computed for both the expected shadow from stellar dynamics and from gas dynamics \citepalias[for details, see][]{M87PaperVI}. The gray band shows the expected range of $\delta$ for the Kerr metric. For the stellar-dynamical prior mass estimates, the EHT measurements show close consistency with the same black hole metric over three orders of magnitude in black hole mass.
}
\label{fig:delta}
\end{center}
\end{figure}

Under the assumption that this range of calibration factors applies generically to all non-Kerr spacetimes that have black hole shadows, we can translate measurements of $\delta$ into constraints on parameters of these spacetimes \citep{Psaltis_2020}. In particular, our measurements exclude specific non-Kerr solutions for \sgra, such as the traversible Morris-Thorne wormhole and the naked singularities in the Reissner-Nordstr\"{o}m metric, which predict shadows that are significantly smaller than those of Kerr black holes.  

Relative to \m87, the primary strengths of testing GR with EHT observations of \sgra are the tight prior constraints on its mass-to-distance ratio $\theta_{\rm g}$ and its shorter gravitational timescale that allows EHT observations to span many dynamical times. Together, these results show SMBHs consistent with predictions of the Kerr metric over a spread of three orders of magnitude in their mass (see \autoref{fig:delta}). The image of \sgra probes a similar gravitational potential to \m87 but spacetime curvature $\xi \propto M^{-2}$ that is six orders of magnitude larger. When combined with constraints from the measurement of gravitational waves from coalescing black-hole binaries with LIGO/Virgo, these results show a striking validation of the predictions of GR over a vast range of scales, from stellar mass black holes to supermassive black holes that are billions of times larger.

\section{Conclusion and Outlook}\label{sec:discuss}

Here we present the results of 2017 EHT observations of \sgra, the central supermassive black hole in the Milky Way. We find evidence for intraday structural variability in \sgra, confirming changes that were hinted at in prior observations across the electromagnetic spectrum. These variations challenge standard approaches to interferometric analysis, so we have developed a variety of methods to infer the structure of this source from our data. From observations on April~7, on which we have the best-sampled data, our analyses consistently reveal a ring-like structure, similar to that seen in \m87. Less complete observations on April~6 support this picture. 

This ring of emission and its central brightness depression closely mirror the structure expected from the plasma in accretion and outflow structures bordering the event horizon of a black hole and partially lensed by its gravity. The angular diameter of this ring is consistent with that expected from the black hole mass inferred from stellar orbits. This consistency allows us to restrict the allowed values of parameters that describe deviations from a Kerr black hole, as predicted by General Relativity. We compare our data, including measurements at other wavelengths, with a large suite of GRMHD simulations. These simulations are remarkably successful at predicting the 1.3\,mm image structure and broadband spectrum of \sgra. However, the 
GRMHD simulations tend to be more variable than the observations, which may be related to our fluid modeling of a collisionless plasma or our neglect of radiative cooling, and only a few configurations can satisfy our full set of observational constraints apart from variability. Our results generally favor models with dynamically strong magnetic fields, moderate (prograde) spin, a lower inclination viewing angle, and strongly decoupled protons and electrons in the emission region. Interestingly, these models also predict a reasonably efficient (compared to the accretion rate) jet outflow, which points to interesting complementary future studies.  However more work is needed to fully explore the physical parameter space, and to understand the variability.  

As the nearest supermassive black hole, \sgra can be scrutinized in ways that are impossible for other sources, making it a unique laboratory for exploring the astrophysics of black holes and testing the predictions of General Relativity. The results presented in these papers are the first EHT contributions to the study of this source, but they are not the last. Subsequent work will characterize the magnetic field configuration of this source through polarimetric observations, as was done for \m87, and describe the structural changes associated with flare activity on April~11. Since 2017, the EHT has continued to gather data using an increasing number of array elements and doubled recording bandwidth. These data will provide improved sensitivity and enable more robust imaging of this dynamic source, eventually allowing movie reconstructions of plasma motions on the $\sim$hour orbital timescales. \\

We thank the anonymous referees for helpful comments that improved the paper. 
The Event Horizon Telescope Collaboration thanks the following
organizations and programs: the Academia Sinica; the Academy
of Finland (projects 274477, 284495, 312496, 315721); the Agencia Nacional de Investigaci\'{o}n 
y Desarrollo (ANID), Chile via NCN$19\_058$ (TITANs) and Fondecyt 1221421, the Alexander
von Humboldt Stiftung; an Alfred P. Sloan Research Fellowship;
Allegro, the European ALMA Regional Centre node in the Netherlands, the NL astronomy
research network NOVA and the astronomy institutes of the University of Amsterdam, Leiden University and Radboud University;
the ALMA North America Development Fund; the Black Hole Initiative, which is funded by grants from the John Templeton Foundation and the Gordon 
and Betty Moore Foundation (although the opinions expressed in this work are those of the author(s) 
and do not necessarily reflect the views of these Foundations);
Chandra DD7-18089X and TM6-17006X; the China Scholarship
Council; China Postdoctoral Science Foundation fellowship (2020M671266); Consejo Nacional de Ciencia y Tecnolog\'{\i}a (CONACYT,
Mexico, projects  U0004-246083, U0004-259839, F0003-272050, M0037-279006, F0003-281692,
104497, 275201, 263356);
the Consejer\'{i}a de Econom\'{i}a, Conocimiento, 
Empresas y Universidad 
of the Junta de Andaluc\'{i}a (grant P18-FR-1769), the Consejo Superior de Investigaciones 
Cient\'{i}ficas (grant 2019AEP112);
the Delaney Family via the Delaney Family John A.
Wheeler Chair at Perimeter Institute; Direcci\'{o}n General
de Asuntos del Personal Acad\'{e}mico-Universidad
Nacional Aut\'{o}noma de M\'{e}xico (DGAPA-UNAM,
projects IN112417 and IN112820); 
the Dutch Organization for Scientific Research (NWO) VICI award
(grant 639.043.513) and grant OCENW.KLEIN.113; the Dutch National Supercomputers, Cartesius and Snellius  
(NWO Grant 2021.013); 
the EACOA Fellowship awarded by the East Asia Core
Observatories Association, which consists of the Academia Sinica Institute of Astronomy and
Astrophysics, the National Astronomical Observatory of Japan, Center for Astronomical Mega-Science,
Chinese Academy of Sciences, and the Korea Astronomy and Space Science Institute; 
the European Research Council (ERC) Synergy
Grant ``BlackHoleCam: Imaging the Event Horizon
of Black Holes" (grant 610058); 
the European Union Horizon 2020
research and innovation programme under grant agreements
RadioNet (No 730562) and 
M2FINDERS (No 101018682);
the Generalitat
Valenciana postdoctoral grant APOSTD/2018/177 and
GenT Program (project CIDEGENT/2018/021); MICINN Research Project PID2019-108995GB-C22;
the European Research Council for advanced grant `JETSET: Launching, propagation and 
emission of relativistic jets from binary mergers and across mass scales' (Grant No. 884631); 
the Institute for Advanced Study; the Istituto Nazionale di Fisica
Nucleare (INFN) sezione di Napoli, iniziative specifiche
TEONGRAV; 
the International Max Planck Research
School for Astronomy and Astrophysics at the
Universities of Bonn and Cologne; 
DFG research grant ``Jet physics on horizon scales and beyond'' (Grant No. FR 4069/2-1);
Joint Princeton/Flatiron and Joint Columbia/Flatiron Postdoctoral Fellowships, 
research at the Flatiron Institute is supported by the Simons Foundation; 
the Japan Ministry of Education, Culture, Sports, Science and Technology (MEXT; grant JPMXP1020200109); the Japanese Government (Monbukagakusho:
MEXT) Scholarship; 
the Japan Society for the Promotion of Science (JSPS) Grant-in-Aid for JSPS
Research Fellowship (JP17J08829); the Joint Institute for Computational Fundamental Science, Japan; the Key Research
Program of Frontier Sciences, Chinese Academy of
Sciences (CAS, grants QYZDJ-SSW-SLH057, QYZDJSSW-SYS008, ZDBS-LY-SLH011); 
the Leverhulme Trust Early Career Research
Fellowship; the Max-Planck-Gesellschaft (MPG);
the Max Planck Partner Group of the MPG and the
CAS; the MEXT/JSPS KAKENHI (grants 18KK0090, JP21H01137,
JP18H03721, JP18K13594, 18K03709, JP19K14761, 18H01245, 25120007); the Malaysian Fundamental Research Grant Scheme (FRGS) FRGS/1/2019/STG02/UM/02/6; the MIT International Science
and Technology Initiatives (MISTI) Funds; 
the Ministry of Science and Technology (MOST) of Taiwan (103-2119-M-001-010-MY2, 105-2112-M-001-025-MY3, 105-2119-M-001-042, 106-2112-M-001-011, 106-2119-M-001-013, 106-2119-M-001-027, 106-2923-M-001-005, 107-2119-M-001-017, 107-2119-M-001-020, 107-2119-M-001-041, 107-2119-M-110-005, 107-2923-M-001-009, 108-2112-M-001-048, 108-2112-M-001-051, 108-2923-M-001-002, 109-2112-M-001-025, 109-2124-M-001-005, 109-2923-M-001-001, 110-2112-M-003-007-MY2, 110-2112-M-001-033, 110-2124-M-001-007, and 110-2923-M-001-001);
the Ministry of Education (MoE) of Taiwan Yushan Young Scholar Program;
the Physics Division, National Center for Theoretical Sciences of Taiwan;
the National Aeronautics and
Space Administration (NASA, Fermi Guest Investigator
grant 80NSSC20K1567, NASA Astrophysics Theory Program grant 80NSSC20K0527, NASA NuSTAR award 
80NSSC20K0645); 
NASA Hubble Fellowship 
grant HST-HF2-51431.001-A awarded 
by the Space Telescope Science Institute, which is operated by the Association of Universities for 
Research in Astronomy, Inc., for NASA, under contract NAS5-26555; 
the National Institute of Natural Sciences (NINS) of Japan; the National
Key Research and Development Program of China
(grant 2016YFA0400704, 2017YFA0402703, 2016YFA0400702); the National
Science Foundation (NSF, grants AST-0096454,
AST-0352953, AST-0521233, AST-0705062, AST-0905844, AST-0922984, AST-1126433, AST-1140030,
DGE-1144085, AST-1207704, AST-1207730, AST-1207752, MRI-1228509, OPP-1248097, AST-1310896, AST-1440254, 
AST-1555365, AST-1614868, AST-1615796, AST-1715061, AST-1716327,  AST-1716536, OISE-1743747, AST-1816420, AST-1935980, AST-2034306); 
NSF Astronomy and Astrophysics Postdoctoral Fellowship (AST-1903847); 
the Natural Science Foundation of China (grants 11650110427, 10625314, 11721303, 11725312, 11873028, 11933007, 11991052, 11991053, 12192220, 12192223); 
the Natural Sciences and Engineering Research Council of
Canada (NSERC, including a Discovery Grant and
the NSERC Alexander Graham Bell Canada Graduate
Scholarships-Doctoral Program); the National Youth
Thousand Talents Program of China; the National Research
Foundation of Korea (the Global PhD Fellowship
Grant: grants NRF-2015H1A2A1033752, the Korea Research Fellowship Program:
NRF-2015H1D3A1066561, Brain Pool Program: 2019H1D3A1A01102564, 
Basic Research Support Grant 2019R1F1A1059721, 2021R1A6A3A01086420, 2022R1C1C1005255); 
Netherlands Research School for Astronomy (NOVA) Virtual Institute of Accretion (VIA) postdoctoral fellowships; 
Onsala Space Observatory (OSO) national infrastructure, for the provisioning
of its facilities/observational support (OSO receives
funding through the Swedish Research Council under
grant 2017-00648);  the Perimeter Institute for Theoretical
Physics (research at Perimeter Institute is supported
by the Government of Canada through the Department
of Innovation, Science and Economic Development
and by the Province of Ontario through the
Ministry of Research, Innovation and Science); the Spanish Ministerio de Ciencia e Innovaci\'{o}n (grants PGC2018-098915-B-C21, AYA2016-80889-P,
PID2019-108995GB-C21, PID2020-117404GB-C21); 
the University of Pretoria for financial aid in the provision of the new 
Cluster Server nodes and SuperMicro (USA) for a SEEDING GRANT approved towards these 
nodes in 2020;
the Shanghai Pilot Program for Basic Research, Chinese Academy of Science, 
Shanghai Branch (JCYJ-SHFY-2021-013);
the State Agency for Research of the Spanish MCIU through
the ``Center of Excellence Severo Ochoa'' award for
the Instituto de Astrof\'{i}sica de Andaluc\'{i}a (SEV-2017-
0709); the Spinoza Prize SPI 78-409; the South African Research Chairs Initiative, through the 
South African Radio Astronomy Observatory (SARAO, grant ID 77948),  which is a facility of the National 
Research Foundation (NRF), an agency of the Department of Science and Innovation (DSI) of South Africa; 
the Toray Science Foundation; Swedish Research Council (VR); 
the US Department
of Energy (USDOE) through the Los Alamos National
Laboratory (operated by Triad National Security,
LLC, for the National Nuclear Security Administration
of the USDOE (Contract 89233218CNA000001); and the YCAA Prize Postdoctoral Fellowship.

We thank
the staff at the participating observatories, correlation
centers, and institutions for their enthusiastic support.
This paper makes use of the following ALMA data:
ADS/JAO.ALMA\#2016.1.01154.V. ALMA is a partnership
of the European Southern Observatory (ESO;
Europe, representing its member states), NSF, and
National Institutes of Natural Sciences of Japan, together
with National Research Council (Canada), Ministry
of Science and Technology (MOST; Taiwan),
Academia Sinica Institute of Astronomy and Astrophysics
(ASIAA; Taiwan), and Korea Astronomy and
Space Science Institute (KASI; Republic of Korea), in
cooperation with the Republic of Chile. The Joint
ALMA Observatory is operated by ESO, Associated
Universities, Inc. (AUI)/NRAO, and the National Astronomical
Observatory of Japan (NAOJ). The NRAO
is a facility of the NSF operated under cooperative agreement
by AUI.
This research used resources of the Oak Ridge Leadership Computing Facility at the Oak Ridge National
Laboratory, which is supported by the Office of Science of the U.S. Department of Energy under Contract
No. DE-AC05-00OR22725. We also thank the Center for Computational Astrophysics, National Astronomical Observatory of Japan.
The computing cluster of Shanghai VLBI correlator supported by the Special Fund 
for Astronomy from the Ministry of Finance in China is acknowledged.

APEX is a collaboration between the
Max-Planck-Institut f{\"u}r Radioastronomie (Germany),
ESO, and the Onsala Space Observatory (Sweden). The
SMA is a joint project between the SAO and ASIAA
and is funded by the Smithsonian Institution and the
Academia Sinica. The JCMT is operated by the East
Asian Observatory on behalf of the NAOJ, ASIAA, and
KASI, as well as the Ministry of Finance of China, Chinese
Academy of Sciences, and the National Key Research and Development
Program (No. 2017YFA0402700) of China
and Natural Science Foundation of China grant 11873028.
Additional funding support for the JCMT is provided by the Science
and Technologies Facility Council (UK) and participating
universities in the UK and Canada. 
The LMT is a project operated by the Instituto Nacional
de Astr\'{o}fisica, \'{O}ptica, y Electr\'{o}nica (Mexico) and the
University of Massachusetts at Amherst (USA). The
IRAM 30-m telescope on Pico Veleta, Spain is operated
by IRAM and supported by CNRS (Centre National de
la Recherche Scientifique, France), MPG (Max-Planck-Gesellschaft, Germany) 
and IGN (Instituto Geogr\'{a}fico
Nacional, Spain). The SMT is operated by the Arizona
Radio Observatory, a part of the Steward Observatory
of the University of Arizona, with financial support of
operations from the State of Arizona and financial support
for instrumentation development from the NSF.
Support for SPT participation in the EHT is provided by the National Science Foundation through award OPP-1852617 
to the University of Chicago. Partial support is also 
provided by the Kavli Institute of Cosmological Physics at the University of Chicago. The SPT hydrogen maser was 
provided on loan from the GLT, courtesy of ASIAA.

This work used the
Extreme Science and Engineering Discovery Environment
(XSEDE), supported by NSF grant ACI-1548562,
and CyVerse, supported by NSF grants DBI-0735191,
DBI-1265383, and DBI-1743442. XSEDE Stampede2 resource
at TACC was allocated through TG-AST170024
and TG-AST080026N. XSEDE JetStream resource at
PTI and TACC was allocated through AST170028.
This research is part of the Frontera computing project at the Texas Advanced 
Computing Center through the Frontera Large-Scale Community Partnerships allocation
AST20023. Frontera is made possible by National Science Foundation award OAC-1818253.
This research was carried out using resources provided by the Open Science Grid, 
which is supported by the National Science Foundation and the U.S. Department of 
Energy Office of Science. 
Additional work used ABACUS2.0, which is part of the eScience center at Southern Denmark University. 
Simulations were also performed on the SuperMUC cluster at the LRZ in Garching, 
on the LOEWE cluster in CSC in Frankfurt, on the HazelHen cluster at the HLRS in Stuttgart, 
and on the Pi2.0 and Siyuan Mark-I at Shanghai Jiao Tong University.
The computer resources of the Finnish IT Center for Science (CSC) and the Finnish Computing 
Competence Infrastructure (FCCI) project are acknowledged. This
research was enabled in part by support provided
by Compute Ontario (http://computeontario.ca), Calcul
Quebec (http://www.calculquebec.ca) and Compute
Canada (http://www.computecanada.ca). 

The EHTC has
received generous donations of FPGA chips from Xilinx
Inc., under the Xilinx University Program. The EHTC
has benefited from technology shared under open-source
license by the Collaboration for Astronomy Signal Processing
and Electronics Research (CASPER). The EHT
project is grateful to T4Science and Microsemi for their
assistance with Hydrogen Masers. This research has
made use of NASA's Astrophysics Data System. We
gratefully acknowledge the support provided by the extended
staff of the ALMA, both from the inception of
the ALMA Phasing Project through the observational
campaigns of 2017 and 2018. We would like to thank
A. Deller and W. Brisken for EHT-specific support with
the use of DiFX. We thank Martin Shepherd for the addition of extra features in the Difmap software 
that were used for the CLEAN imaging results presented in this paper.
We acknowledge the significance that
Maunakea, where the SMA and JCMT EHT stations
are located, has for the indigenous Hawaiian people.


\facility{EHT, ALMA, APEX, IRAM:30m, JCMT, LMT, SMA, ARO:SMT, SPT, Chandra, EAVN, GMVA, NuSTAR, Swift, VLT}.

\bibliography{bib,EHTCPapers}

\allauthors

\end{document}